\documentclass[a4paper,12pt]{article}
\usepackage{fullpage,epsf}
\usepackage{amssymb,amsmath,amsthm}
\usepackage[pdftex]{graphicx}
\def\ltsima{$\; \buildrel < \over \sim \;$}
\def\lsim{\lower.5ex\hbox{\ltsima}}
\def\gtsima{$\; \buildrel > \over \sim \;$}
\def\gsim{\lower.5ex\hbox{\gtsima}}
\numberwithin{equation}{section}
\begin{document}
\pagestyle{empty}                       
\epsfxsize=40mm                         
\hfill
\begin{minipage}[b]{110mm}
        \center{
        \baselinestretch \linespread{3}
        {\Huge\bf School of Physics and Astronomy
        \vspace*{17mm}}}
\end{minipage}
\hfill
\begin{minipage}[t]{40mm}
        \includegraphics[width=35mm]{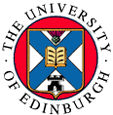}
\end{minipage}
\par\noindent                                           
\vspace*{2cm}
\begin{center}
        {\Large Institute for Astronomy}\\
        {\Large\bf \Large\bf Edinburgh Centre for Computational Astrophysics}\footnote{http://www.roe.ac.uk/$\sim$aam/ecca}\\
\vspace*{0.5cm}
        {\LARGE\bf {\texttt{IGMtransmission}}: A Java GUI to model the}\\
\vspace*{0.2cm}
{\LARGE\bf effects of the Intergalactic Medium on the colours}\\
\vspace*{0.15cm}
{\LARGE\bf of high redshift galaxies}        
\end{center}
\vspace*{0.5cm}
\begin{center}
        \bf Christopher M. Harrison, Avery Meiksin, David Stock\\                           
        31st May 2011                                    
\end{center}
\vspace*{5mm}
%
%
\begin{abstract}
{\texttt{IGMtransmission}}\footnote{{\texttt{IGMtransmission}} is copyrighted, and is licensed under the GNU General Public License, Version 3 (http://www.gnu.org/licenses/gpl.html). It is available for download from http://code.google.com/p/igmtransmission} is a Java graphical user interface that implements Monte Carlo simulations to compute the corrections to colours of high-redshift galaxies due to intergalactic attenuation based on current models of the Intergalactic Medium. The effects of absorption due to neutral hydrogen are considered, with particular attention to the stochastic effects of Lyman Limit Systems.
\end{abstract}

\vspace*{1cm}



\vspace*{2cm}

\newpage
%
\pagestyle{plain}                               
\setcounter{page}{1}                            
\tableofcontents                                
\newpage

\section{Introduction}\label{SEC:introduction}
{\texttt{IGMtransmission}} is a Java graphical user interface (GUI) that models the effect of intergalactic neutral hydrogen attenuation on the colours of objects based on the model of Meiksin(2006). The code allows the effects to be modelled for a range of galaxies and filters provided. The galaxy spectra are taken from the stellar $+$ nebular emission line models of Leitherer et al. (1999). Photometric filters are included for the {\it Hubble Space Telescope}\footnote{http://hubblesite.org}, the Keck telescope\footnote{http://www.keckobservatory.org}, the Mt. Palomar 200-inch\footnote{http://www.astro.caltech.edu/palomar}, the SUBARU telescope\footnote{http://www.naoj.org} and UKIDSS\footnote{http://www.jwst.nasa.gov}. Alternative spectra and filters may be straightforwardly added by the user.

To model the colours of high-redshift galaxies a model galaxy spectrum is taken, shifted appropriately for a chosen redshift, and then attenuated to account for intervening neutral hydrogen using then chosen model and set of parameters. Monte Carlo simulations over many lines of sight are performed, with the option of obtaining each colour independently or calculating an average. To do this, a population of discrete, Lyman Limit Systems are drawn from the distributions, $dN/dz$ and $dN/d\tau_L$, outlined in Section \ref{SEC:models}, placed along random lines of sight and their total contribution to the intergalactic transmission function calculated. In addition to the default choice, the alternative Lyman Limit Systems distribution of Inoue \& Iwata (2008) is allowed for. Alternatively, the parameter values defining the model may be chosen by the user. Specifying a fixed set of individual Lyman Limit Systems is also accommodated. Mean contributions to the opacity due the Ly$\alpha$ forest and the photoelectric absorption of the optically thin IGM are then included. The colours and IGM k-corrections are computed through the chosen broad-band filters, given this amount of attenuation for many lines of sight. The cumulative probability distributions, over the different lines of sight, are also produced. The model spectra, broad-band filter combinations, attenuation model parameters and output options are all chosen via the GUI.

The code requires Java v.1.6. It has been tested on {\it Windows   Vista\texttrademark Home Basic}, Debian Linux v5.0 and OS X v.10.5.8. The code uses the random number generator included in the Sun Microsystems' Java Development Kit v.1.6. To query the default version of Java running on a given system, type ``java -version.'' If it is not v.1.6, assistance from a system administrator may be required to obtain it. A screenshot of the GUI and an explanation of how it is used is included in Appendix \ref{GUI}. The details of the code design are provided in the next section, followed by a section describing the various code features.

\section{Code design}\label{SEC:design}

\subsection{Modelling intergalactic attenuation}\label{SEC:models}

Intervening absorption systems are likely to be of various origins; for example the outer edge of galaxies, halo gas and diffuse material in intergalactic space. The absorbers are generally considered to be discrete and separate systems. They are classified into three main categories depending on their properties:\ Ly$\alpha$ forest absorbers (with column densities $N_{\rm HI} \lsim 10^{17} {\rm cm}^{-2}$), Lyman Limit Systems (LLSs) ($10^{17} {\rm cm}^{-2} \lsim N_{\rm HI} \lsim 10^{20} {\rm cm}^{-2}$) and Damped Ly$\alpha$ Absorbers (DLAs) ($N_{\rm HI} \gsim 10^{20}{\rm cm}^{-2}$) (Meiksin 2009). It has been suggested that there is an upper-cut off at $3-5\times10^{21}\,{\rm cm}^{-2}$, and systems with column densities lower than $10^{12}\, {\rm cm}^{-2}$ may exist but are difficult to detect. The Ly$\alpha$ forest absorbers are by far the most common, and the DLAs the rarest and most easily identifiable in spectra. The intermediate LLSs have a sufficiently large column density to absorb photons with energies above the photoelectric edge (the Lyman limit). The classifications are not strictly exclusive; for example, DLAs will also produce LLSs. The total effective optical depth of the IGM is the sum of the optical depths due to each absorbing system. 
The two main effects responsible for attenuation of the spectrum of a distant light source are the photoelectric absorption by LLSs and optically thin IGM and resonance line scattering by the Ly$\alpha$ forest. Additional contributions are made by intervening metal systems and helium. Contributions of metals and HeI (neutral helium) are small (Madau 1995), while HeII (singly ionised helium) will contribute only at wavelengths  $\lambda < 228(1+z)$ \AA. This is outside the wavelength range considered. Due to the discrete and clumpy nature of these systems, it is not possible to know the amount of absorption along one particular line of sight. However, it is possible to predict a mean amount of absorption and model the statistical fluctuations. Madau (1995) uses analytical approximations to derive the mean attenuation. More recently work has been done using Monte Carlo simulations to predict the mean attenuation for the models used in the code.

\subsubsection*{Lyman transitions}
The Ly$\alpha$ forest does not produce significant fluctuations, therefore a
mean contribution is assumed. The mean optical depth for the Lyman transition $n\rightarrow 1$ is defined by
\begin{equation}\label{meanopticaldepth}
\bar{\tau}_n \equiv -\ln\langle \exp(-\tau_n)\rangle
\end{equation}
where $\langle \exp(-\tau_n) \rangle$ is the corresponding mean transmitted flux and $n$ corresponds to $\alpha$, $\beta$ etc. The values of $\tau_n$ used in this investigation for transitions up to $n=31$ have been calculated based on the simulation of a cold dark matter model and are given by Meiksin(2006). Inoue \& Iwata (2008) adopt a different technique for calculating the absorption due to the Ly$\alpha$ forest. However, both methods agree excellently for the intergalactic transmission at wavelengths greater than the Lyman limit (see Figure 6 in Inoue \& Iwata 2008). Therefore, the treatment as prescribed by Meiksin(2006) is assumed throughout this investigation. Both models are based on recent estimates for the properties of the Ly$\alpha$ forest, which show substantial differences from Madau (1995).

\subsubsection*{Photoelectric absorption}
\begin{figure}[hbtp]
\begin{center}
\includegraphics[width=0.4\textwidth]{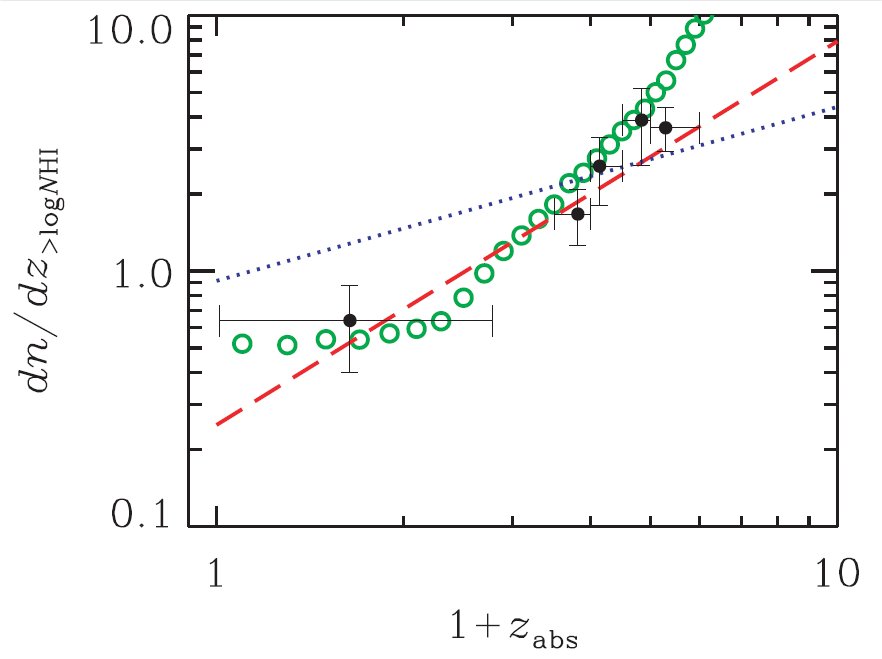}
\end{center}
\caption{{\small The number of LLSs per unit red shift along an average line of sight as a function of the LLSs' redshift. The filled circles are the recent observed data by P\'eroux et al.(2005). The open circles are the model assumed by Inoue \& Iwata (2008). The dashed line is the assumed distribution in Meiksin (2006). The dotted line is that assumed by Madau (1995).}}
\label{FIG:zDistributions}
\end{figure}

The contribution to the optical depth by photoelectric absorption for a source at redshift $z$ is split into two parts (Meiksin 2006):\ the contribution from systems optically thin ($\tau_L\leq1$) at the Lyman limit and the contributions from the LLSs. The contribution from optically thin systems is revised from Meiksin (2006) according to
\begin{equation}\label{diffuseIGM}
\tau _{L}^{\rm IGM} = A(1+z_L)^{4.4}\left[\frac{1}{(1+z_L)^\frac{3}{2}}-\frac{1}{(1+z)^\frac{3}{2}}\right]
\end{equation}
where $z_L=\lambda /\lambda _{L}-1$ and $\lambda_{L} = 912$ \AA. The normalisation is taken to be $A=0.07553$, which is based on simulations. There is a degree of uncertainty in the amount of the absorption, depending on the ionization state of the hydrogen in the IGM and on the cosmological parameters assumed. The contribution due to the optically thick, $\tau_L>1$, Lyman Limit Systems is given by
\begin{equation}\label{LLSattenuation}
\tau_{L}^{\rm LLS} = \int^{z}_{z_{L}} dz' \int^{\infty}_{1} d\tau_{L}\frac{\partial^2 N}{\partial \tau_L \partial z'}\left\{1-\exp\left[-\tau_L\left(\frac{1+z_L}{1+z'}\right)^{3} \right] \right \}
\end{equation}
where $\frac{\partial^2 N}{\partial \tau_L \partial z'}$ is the number of absorbers along the line of sight per unit redshift interval per unit optical depth of the system. The spatial distribution of LLSs adopted in Meiksin (2006) is
\begin{equation}\label{MeiksinDistribution}
\frac{dN}{dz} = N_0(1+z)^{\gamma} 
\end{equation}
where $N_0 = 0.25$ and $\gamma = 1.5$. The optical depth distribution is given by
\begin{equation}\label{TauDistribution}
\frac{dN}{d\tau_L} \propto  \tau_L^{-\beta}  
\end{equation}
where $\beta = 1.5$, although there may be small deviations from a perfect power law (Meiksin 2009).
\newline\newline
Inoue \& Iwata (2008) treat the photoelectric contribution to the optical depth differently, as shown in Figure 6 of their paper. At wavelengths shorter than the Lyman limit there is some disagreement from the model used in Meiksin (2006). Inoue \& Iwata treat the DLAs and LLSs separately, but they show that the contribution of DLAs to the Lyman continuum is small. They assume the distribution for $\tau_L$ given by Eq. \eqref{TauDistribution} for the LLSs using $\beta = 1.3$, based on a fit by eye to recent observational data. A common $z$ distribution for all types of absorbers is assumed, but this distribution normalised to only include the optically thick LLSs is given by
\begin{equation}\label{InoueDistribution}
\frac{dN}{dz} = \frac{A}{688.4} \left\{
\begin{array}{ll}
\left(\frac{1+z}{1+z_1}\right)^{\gamma_1}  & (0 < z \leq z_1) \\
\left(\frac{1+z}{1+z_1}\right)^{\gamma_2}  & (z_1 < z \leq z_2) \\
\left(\frac{1+z_2}{1+z_1}\right)^{\gamma_2}\left(\frac{1+z}{1+z_2}\right)^{\gamma_3} & (z_2 < z) \\
\end{array}
\right.
\end{equation}
where they adopt $z_1 = 1.2$, $\gamma_1 = 0.2$, $\gamma_2 = 2.5$, $z_2 = 4.0$, $\gamma_3 = 4.0$ and $A = 400$. Figure \ref{FIG:zDistributions}, taken from Inoue \& Iwata (2008), shows a comparison between the $z$ distributions for the LLSs adopted by Meiksin (2006), Inoue \& Iwata (2008), Madau (1995) and some recent observational data. It can be seen there is good agreement between the observed data and the distributions from the two more recent models. If the Inoue \& Iwata (2008) option is selected, the code uses only their LLS distribution while retaining the model of Meiksin (2006) for the optically thin IGM and the Ly$\alpha$ forest.

\subsubsection*{The intergalactic transmission function}
\begin{figure}[hbtp]
\begin{center}
\includegraphics[width=0.5\textwidth]{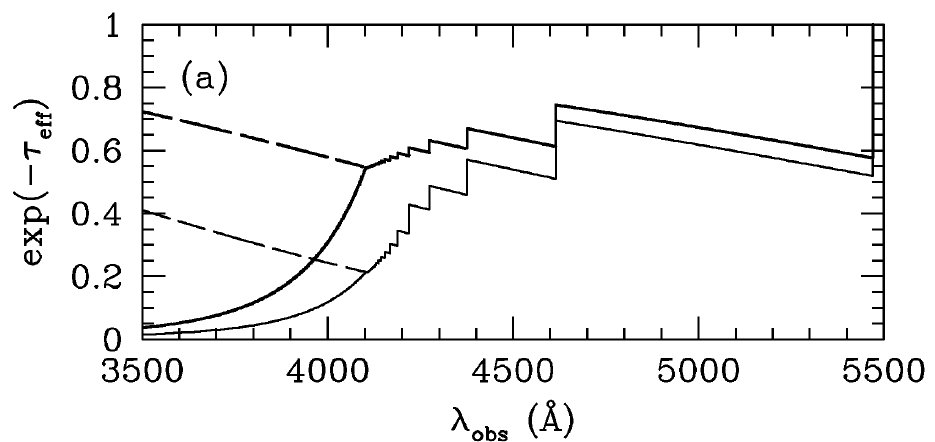}
\end{center}
\caption{{\small Intergalactic transmission as a function of observed wavelength. Mean intergalactic transmission for a source at redshift $z = 3.5$ (solid lines). The mean intergalactic transmission assuming no LLSs lie along the line of sight is shown by the dashed lines. (N.B.\ This uses an older model for the contribution to the optically thin IGM.) The thick lines are those using the model assumed by Meiksin(2006); the light lines are the estimates of Madau(1995).}}
\label{FIG:transmissionExample}
\end{figure}

The total effective optical depth of the IGM,  $\tau_{\rm eff}$, is the sum of the resonance line and photoelectric contributions. It is a function of the observed wavelength, as well as the redshift of the source, and is given by 
\begin{equation}
\tau_{\rm eff} = \tau_{\rm LC} + \tau_{\rm LC}^{\rm IGM} + \sum_i \tau_{i}
\end{equation}
where $\tau_{\rm LC}$ is the total Lyman continuum contribution from all LLSs, $\tau_{\rm LC}^{\rm IGM}$ is the contribution from the optically thin IGM, Eq. \eqref{diffuseIGM}, and $\tau_{i}$ is the optical depth for the $i$th Lyman series line. The attenuation of a background source by the IGM may then be characterised by the transmission function $\exp(-\tau_{\rm eff})$. An example of the mean transmission through the IGM, taken from Meiksin (2006), is given in Figure \ref{FIG:transmissionExample}. A value of 1 represents complete transmission through the IGM, at an observed wavelength $\lambda_{\rm obs}$, while a value of 0 represents no transmission (hence the source is blacked out at these wavelengths). Each step observed in the transmission function, in the region of wavelengths greater than the Lyman Limit ($\lambda_{\rm obs} > (1+z)912$ \AA), represents the addition of a new line in the Lyman series. The Lyman continuum contributes at wavelengths shorter than the Lyman limit.

\subsection{Broad band magnitudes, colours and IGM k-corrections}\label{SEC:colours}

The apparent {\it AB-magnitude} of a source of intrinsic flux $f_{\lambda}$ (in CGS units) measured through a filter with transmission $T(\lambda)$ and effective optical depth of attenuation, $\tau_{\rm eff}$, (see previous section) is given by
\begin{equation}\label{ABmag}
m_{\rm AB} = -2.5\log_{10}\int{d\log\lambda\, f_{\lambda} (\lambda^2/c) e^{-\tau_{\rm eff}}T(\lambda)} - 48.59
\end{equation}
(Fukugita et al. 1996),
where $\lambda$ is the wavelength. The code uses the normalisation convention for $T(\lambda)$ of
\begin{equation}\label{Tnormalization}
\int_0^{\infty} d\log\lambda\, T(\lambda) = 1.
\end{equation}\newline\newline

Alternative magnitudes on the {\it Vega-magnitude} system, for which the star Vega is defined as having zero magnitude in all bands, are also allowed for.

The difference between the magnitudes with and without intergalactic attenuation is defined as the IGM {\it k}-correction
\begin{equation}\label{kCorrection}
k_{\rm IGM} = m_{\rm AB}(\tau_{\rm eff}) - m_{\rm AB}(\tau_{\rm eff}=0), 
\end{equation}
where $m_{\rm AB}(\tau_{\rm eff})$ is the magnitude through the IGM with effective optical depth $\tau_{\rm eff}$, and $m_{\rm AB}(\tau_{\rm eff}=0)$ is the magnitude if there were no attenuation due to the IGM.

\section{Code features}

\subsection{Calculating intergalactic transmission functions}\label{SEC:methodTransmission}
The code is designed to reproduce the intergalactic transmission curves in Meiksin (2006) by using Monte Carlo simulations. The average number $\bar{N}$ of LLSs is calculated by integrating the assumed distribution, Eq. \eqref{MeiksinDistribution}, between $z=0$ and $z$ (the chosen redshift of the source). The number of LLSs actually intercepted along each line of sight is then drawn from the Poisson distribution, $f(k,\bar{N}) = \bar{N}^k e^{-\bar{N}}/k!$, where $k$ is the number of systems. For $k>10$, the limiting Gaussian distribution is used. For each LLS, a value of $z$ and $\tau_{L}$ is assigned drawn randomly from the distributions Eq. \eqref{MeiksinDistribution} and Eq. \eqref{TauDistribution}. The contribution to attenuation by the diffuse IGM, Eq.\eqref{diffuseIGM}, is also included. The normalisation $A$ is an input parameter in the GUI.
\newline\newline
For each LLS, with optical depth $\tau_L$ and redshift $z$, the photoelectric contribution to the optical depth at a given wavelength $\lambda$ is calculated using
\begin{equation}
\tau_{\rm LC}(\lambda) = -\tau_{L}\left[\frac{\lambda}{\lambda_L(1+z)}\right]^3
\end{equation}
which is derived from Eq. \eqref{LLSattenuation} and $\lambda_L = 912$ \AA. This is calculated over a certain wavelength range with a given resolution. The default resolution is $1$ \AA, but this may be altered by the user. The contributions from each LLS along the line of sight are then summed. The mean contribution due to each line in the Lyman series as derived in Meiksin (2006) are summed on top of the photoelectric contribution.
\newline\newline
The process is repeated for each realisation (i.e., line of sight) and an average is then taken to get the total effective optical depth of the IGM and hence transmission function $\exp(-\tau_{\rm eff})$ (see Section \ref{SEC:models}).\newline\newline

An option is included to distribute the redshifts of the LLSs according to that of Inoue \& Iwata (2008), Eq. \eqref{InoueDistribution}. As discussed in Section \ref{SEC:models}, Meiksin (2006) and Inoue \& Iwata (2008) agree closely on the resonance line scattering contribution to the IGM transmission. The user may use this option to examine the effect of distributing the LLSs according to the empirically derived $z$ distribution of Inoue \& Iwata, although the model used by Meiksin (2006) for resonance line scattering and the optically thin IGM (Eq. \eqref{diffuseIGM}) is retained. This option is implemented using the same method as for the Meiksin (2006) distribution. The parameters of the distribution function are included as input options for generality; these may be adjusted by clicking on the appropriate tab, which enables the user to enter alternative values.

It is also possible to compute the effect of a fixed set of LLSs on the
galaxy colours instead of using a Monte Carlo realisation. This may be done
by adding a file in the top-level directory (such as input.dat) containing
in the first line the number of LLSs, followed by a line for each absorber containing the data pair $z$ and $\tau_L$ (with a space between), corresponding to the redshift and Lyman limit optical depth of the absorber.

\subsection{Calculating broad-band magnitudes, colours and IGM k-corrections}\label{SEC:calculating}
The code computes the broad-band magnitudes and colours of model galaxy spectra for a choice of broad-band filters.

\subsubsection*{Filters}
A large choice of filter transmission curves (see Section \ref{SEC:colours}) were included and can be selected from drop-down menus in the GUI.\footnote{This catalogue can easily be extended by the user, if a particular survey is to be investigated, as explained in Section \ref{SEC:calculating}.} The following filters have been implemented:
\begin{itemize}
\item {\it Hubble Space Telescope (HST)}\footnote{http://www.stsci.edu/hst}:\ the filter responses for F300W, F450W and F606W for the Wide Field Planetary Camera 2 (WFPC2), referred to as $U_{300}$, $B_{450}$ and $V_{606}$; the filter responses for F435W, F606W, F775W and F850LP for the Advanced Camera for Surveys (ACS), referred to as $B_{435}$ (approximating Johnson B), $V_{606}$, $i_{775}$ (approximating SDSS $i$) and $Z_{850}$ (approximating SDSS $z$); the filter responses for F110W and F160W for the Near Infrared Camera and Multi-Object Spectrometer (NICMOS), referred to as $J_{110}$ and $H_{160}$.
\item {\it UK Infrared Telescope (UKIRT)}:\ the filter responses for the UKIRT Infrared Deep Sky Survey (UKIDSS) filters $H$, $J$, $K$, $Y$ and $Z$; the filter responses for the Wide Field Camera (WFCAM) filters $J$ and $K$.
\item {\it Subaru XMM-Newton Survey (SXDS)}\footnote{Provided by M. Cirasuolo (private communication).}: the filter responses for $B$, $V$, $R$, $i^\prime$ and $z^\prime$.
\item {\it Keck}:\ the filter responses for the Low Resolution Imaging Spectrometer (LRIS) $B$, $V$, $R_s$, $R_{\rm KC}$ (Kron-Cousins) and $V$.
\item {\it Palomar 200-inch}:\ the filter responses for $U_n$, $G$ and $R$.
\item {\it Canada-France-Hawaii Telescope (CFHT)}:\ The filter response for $u$.
\end{itemize}

The response curves also include estimates for the total system throughput (i.e., including the response of the full system). Filter transmission curves $T(\lambda)$ are normalized according to Eq. \eqref{Tnormalization}. Gaussian quadrature (see Appendix \ref{GaussianQuadrature}) is used to compute the normalisation constant.

The user should verify each filter response against expectation before performing any quantitative analysis. Additional filter response curves may be readily incorporated by adding files containing them to the {\texttt{filters}} directory with the following format:\ the first line should contain $len$ and $n$ (separated by a space), followed by $len$ pairs of $\lambda$ and $T(\lambda)$ (separated by a space), where $len$ is the number of wavelengths, $n$ is a multiplyer of the wavelength $\lambda$ such that $n\lambda$ is the wavelength in \AA, and $T(\lambda)$ is the filter transmission at wavelength $\lambda$.

\subsubsection*{Spectra}
A variety of different intrinsic galaxy spectra are available for selection, taken from the {\texttt{STARBURST99}} model of Leitherer et al. (1999). The GUI is designed so that an age and metallicity of the model galaxy may be chosen, as well as an option to chose whether to use an instantaneous or continuous star formation model. The models use a Salpeter Initial Mass Function\footnote{The Salpeter Initial Mass Function is given by $\xi(M)=\xi_0M^{-2.35}$, where $\xi (M)dM$ is the number of stars born with masses between $M$ and $M+dM$ and $\xi_0$ is a normalisation constant.} with stellar masses in the range $1<M/M_{\odot}<100$. The models include emission from both stars and nebulae. The spectra may be chosen from a drop down menu and radio buttons. The $f_{\lambda}$ values are linearly interpolated to the same wavelength values as the normalized transmission functions $T(\lambda)$ of the filters. The choice of redshift $z$ for the model galaxy is included as an input parameter. The wavelengths of the spectra are consequently shifted by $(1+z)$. Internal reddening by gas and dust within the galaxies is not included.

Additional spectra may be readily incorporated by adding files containing them to the {\texttt{uploadedSpectra}} directory with the following format:\ the first line should contain $len$ and $n$ (separated by a space), followed by $len$ pairs of $\lambda$ and $f_\lambda$ (separated by a space), where $len$ is the number of wavelengths, $n$ is a multiplyer of the wavelength $\lambda$ such that $n\lambda$ is the wavelength in \AA, and $f_\lambda$ is the wavelength-specific flux at wavelength $\lambda$.

\subsubsection*{Calculating magnitudes and colours}
The transmission curves $\exp(-\tau_{\rm eff})$ are calculated using the methods outlined in Section \ref{SEC:methodTransmission} and then linearly interpolated to the same wavelength values as the filter transmission curves $T(\lambda)$. The interpolated values of $\exp(-\tau_{\rm eff})$, $T(\lambda)$ and redshifted $f_{\lambda}$ are then multiplied together and the AB-magnitudes calculated using the integral \eqref{ABmag}, again with Gaussian quadrature. An option is provided to compute magnitudes referenced to Vega.

\subsubsection*{IGM k-corrections}
The IGM k-correction, Eq. \eqref{kCorrection}, for each filter is computed by also calculating the AB-magnitude 
 with no intergalactic attenuation (i.e., $\tau_{\rm eff} = 0$) and subtracting this from the magnitude including attenuation.
 
\subsection{Distribution of absorbers}
\begin{figure}[!ht]
\begin{center}
\includegraphics[width=0.7\textwidth]{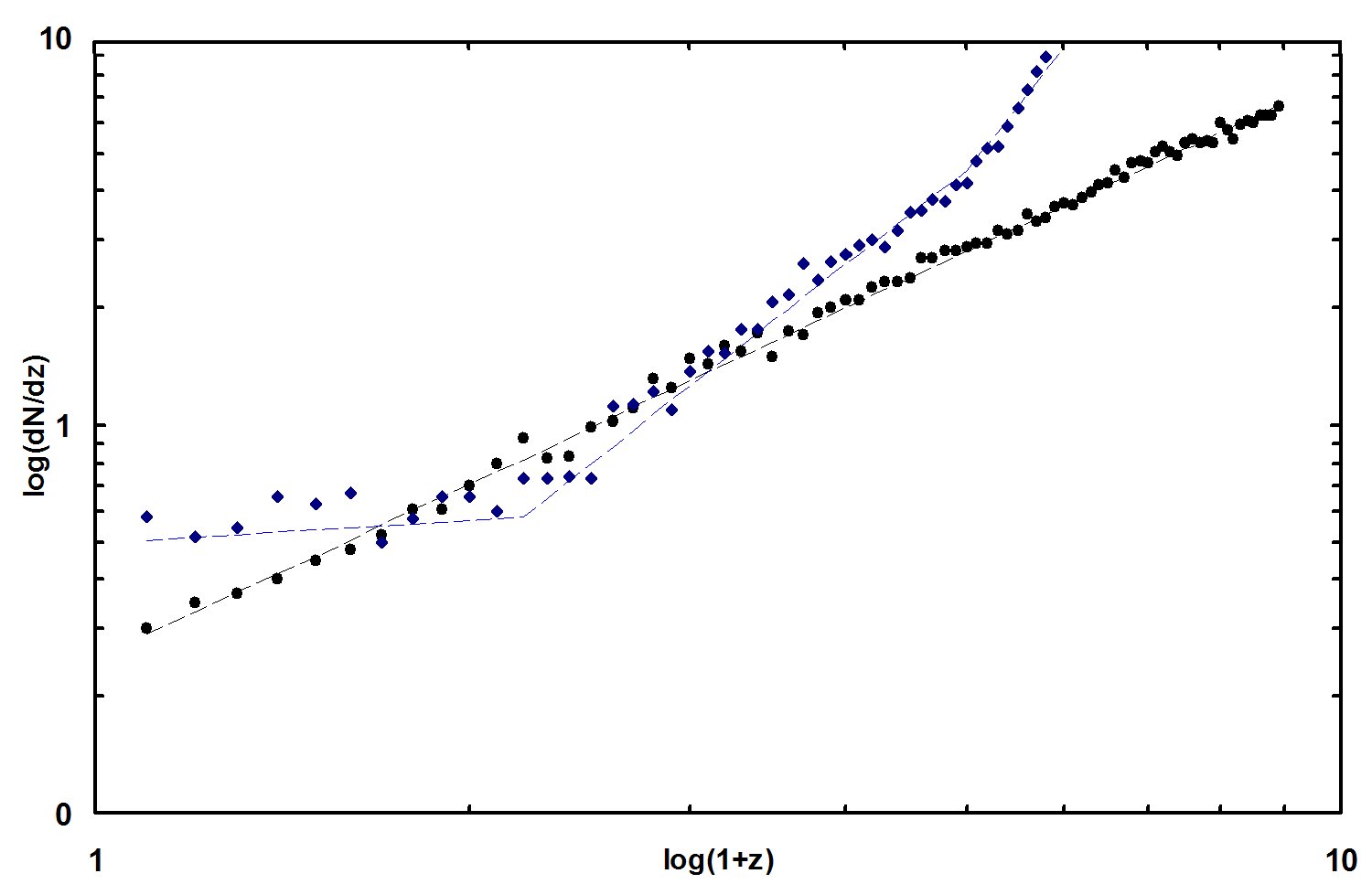}
\end{center}
\caption{\small Number of LLSs per unit redshift along an average line of sight. The black points represent the output of the Monte Carlo simulation using the Meiksin(2006) model with 2000 realisations. The black dashed line shows the expected result using Eq. \eqref{MeiksinDistribution}. The blue diamonds represent the output of the Monte Carlo simulation using the Inoue \& Iwata (2008) model. The blue dashed line shows the expected results using Eq. \eqref{InoueDistribution}.}
\label{FIG:distributionFunctions}
\end{figure}

Figure \ref{FIG:distributionFunctions} shows the resultant $z$ distribution of LLSs as produced by the Monte Carlo simulation. This test was done by running the simulation for a galaxy at redshift, $z = 8$, for 2000 realisations and outputting the properties of each LLS. These were put into redshift bins of width, $\Delta z$, and then plotted using the approximation $dN/dz \sim \Delta N/\Delta z$ (averaged over all realisations), where $\Delta N$ is the number of absorbers produced in each bin. The figure shows the distribution of LLSs according to both Meiksin (2006) (Eq. \eqref{MeiksinDistribution}) and Inoue \& Iwata (2008) (Eq. \eqref{InoueDistribution}), using the parameters suggested in their respective papers. The results of the Monte Carlo simulations are seen to match the expected distributions. The data are fairly noisy at the lower redshifts due to the small number of LLSs in each redshift bin.

\subsection{Calculating the mean intergalactic transmission functions}\label{SEC:resultsTransmission}
\begin{figure}[!ht]
\begin{center}
\includegraphics[width=0.7\textwidth]{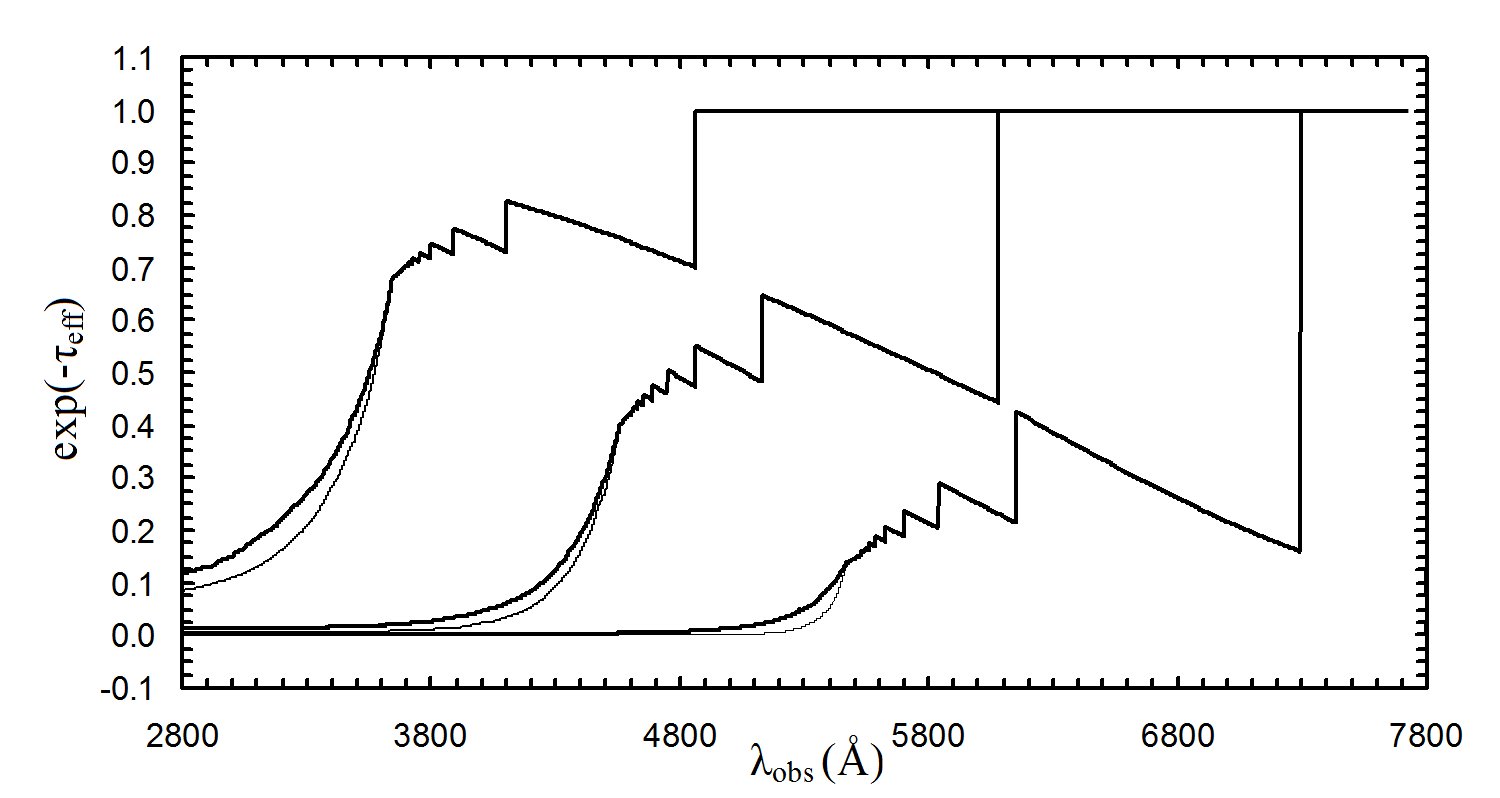}
\end{center}
\caption{\small{Mean intergalactic transmission, over 2000 lines of sight, as a function of observed wavelength, $\lambda_{\rm obs}$, for sources at $z = $ 3, 4 and 5 as viewed from left-to-right. The heavy lines assume the distribution of LLSs given in Eq. \eqref{MeiksinDistribution}, the light lines assume that given in Eq. \eqref{InoueDistribution}. The values of $\beta$ in Eq. \eqref{TauDistribution} used are 1.5 and 1.3, respectively.}} 
\label{FIG:RESULTtransmission}
\end{figure}

Example mean transmission curves as produced by the Monte Carlo simulation are shown in Figure \ref{FIG:RESULTtransmission}. These are calculated assuming the model for the Ly$\alpha$ forest as given in Meiksin (2006) and the contribution due to the optically thin IGM, according to Eq. \eqref{diffuseIGM}. The result of distributing the LLSs according to both Eq. \eqref{MeiksinDistribution} and Eq. \eqref{InoueDistribution} are shown for comparison. The steps above the Lyman limit due to the Lyman transitions are observed. The Lyman continuum below the Lyman limit, due to the LLSs and optically thin IGM, can clearly be seen. The transmission in the continuum regime according to Inoue \& Iwata (2008) is smaller because of the generally larger number of LLSs along each line of sight (as demonstrated in Figure \ref{FIG:distributionFunctions}).

\subsection{Transmission along individual lines of sight}
\begin{figure}[!ht]
\begin{center}
\includegraphics[width=0.5\textwidth]{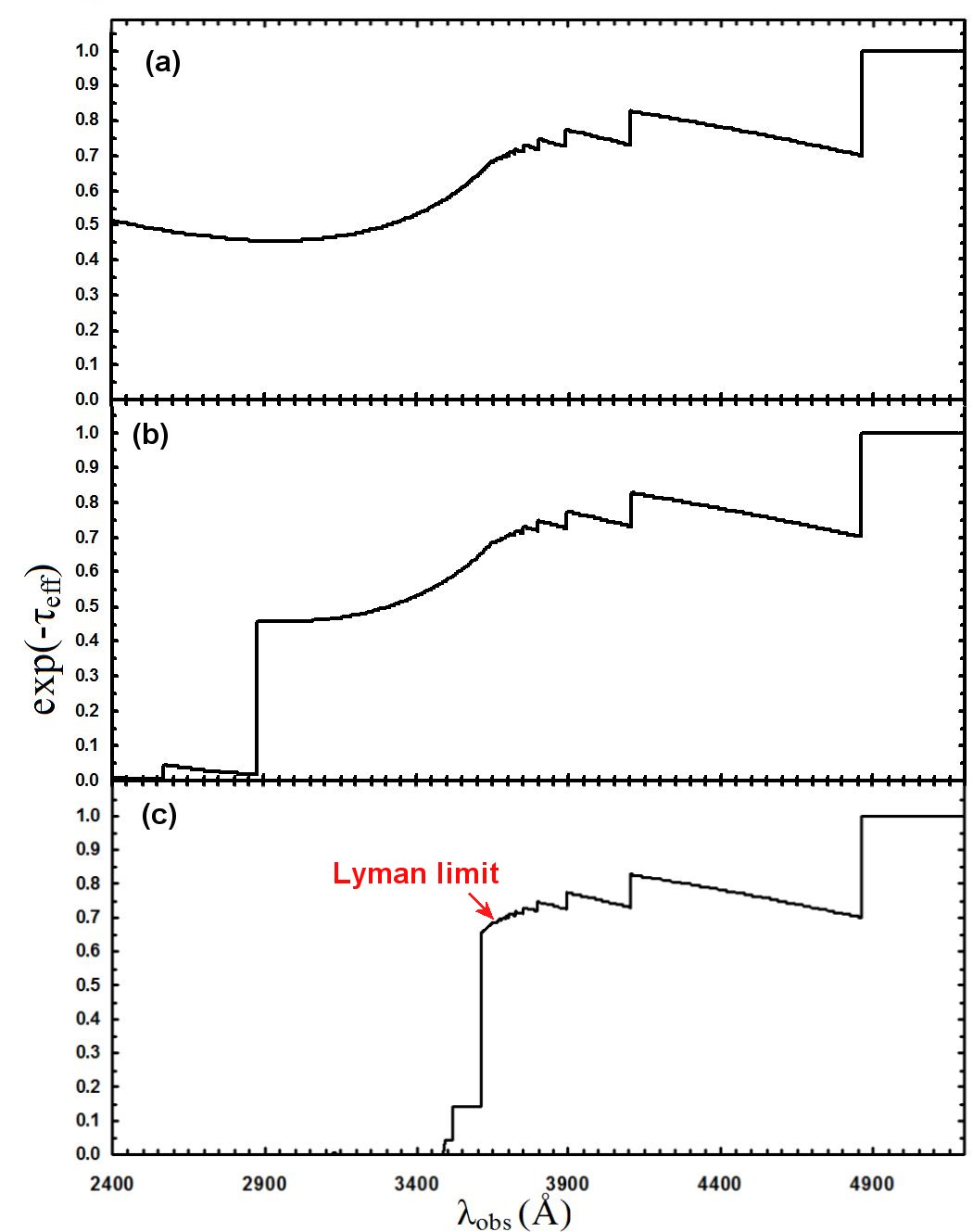}
\end{center}
\caption{\small{Example transmission curves for a source at $z=3$ along three separate lines of sight, assuming the $z$ distribution of LLSs Eq. \eqref{MeiksinDistribution} and $\beta=1.5$. Panel (a) shows the transmission when there are no LLSs near the source; (c) when there are a several absorbers near the source and (b) is an intermediate case.}}
\label{FIG:randomLinesOfSight}
\end{figure}

There is a stochastic nature of the amount of attenuation along different lines of sight due to the rare LLSs. The result is that the transmission in the Lyman continuum regime ($\lambda_{\rm obs} < (1+z)912$ \AA) can vary dramatically along different lines of sight. A mean contribution from the Ly$\alpha$ forest is assumed (see Section \ref{SEC:models}). Figure \ref{FIG:randomLinesOfSight} demonstrates the transmission for three separate lines of sight for a source at $z=3$. In the two extreme cases, a spectrum could be completely truncated below the Lyman limit, or if there are no LLSs near the source, significant transmission could still occur. Indeed, for a source at redshift $z = 3$, assuming Eq. \eqref{MeiksinDistribution}, the average number of intervening LLSs expected is 3.1. Therefore, from the Poisson distribution, the probability that no LLSs occur at all is $e^{-3.1}=4.5\%$.

\section{Acknowledging use of the code}

The use of {\texttt{IGMtransmission}} should be acknowledged using a statement like

\begin{minipage}[b]{160mm}
        \baselinestretch \linespread{3}
This paper computed IGM transmission values using {\texttt{ IGMtransmission}}\footnote{Available for download from http://code.google.com/p/igmtransmission} (Harrison, Meiksin \& Stock 2011), based on the transmission curves of Meiksin (2006).
\end{minipage}

If the LLS distribution from Inoue \& Iwata (2008) is used, a reference to their work should be added as well. Any work using the model galaxy spectra should reference Leitherer et al. (1999).


\section*{Appendices}
\appendix
\section{The Graphical User Interface}\label{GUI}
\begin{figure}[!ht]
\begin{center}
\includegraphics[width = 0.5\textwidth]{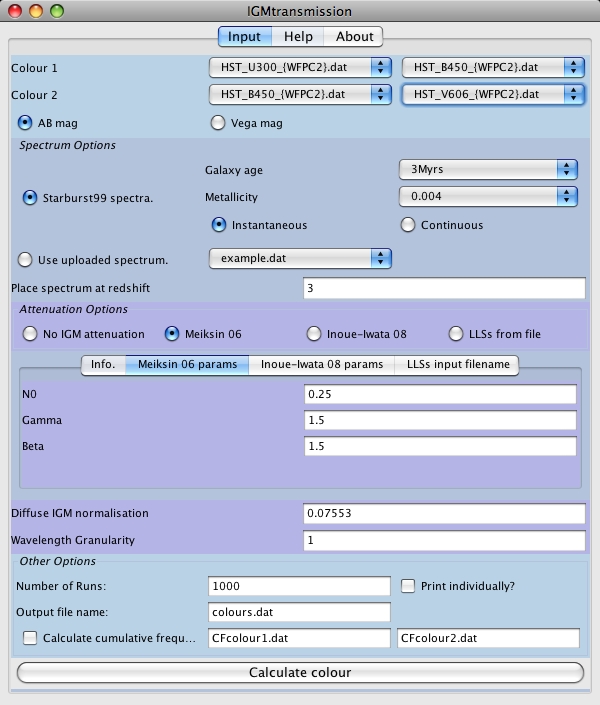}
\end{center}
\caption{The GUI produced to allow easy input of the numerous options and parameters associated with the program.}
\label{FIG:GUI}
\end{figure}
Filter transmission functions may be chosen from the drop down menus. The program calculates two colours simultaneously; for example, if the program was in the set-up shown in Figure \ref{FIG:GUI} it will calculate colours $U_{300}-B_{450}$ and $B_{450}-V_{606}$ for the Hubble Space Telescope, Wide Field Planetary Camera 2 (WFPC2) including estimates for the total system throughput. These are taken from a library of transmission curves in a directory associated with the GUI. This catalogue is easily extended by simply adding transmission curve files to the directory, as described in Section \ref{SEC:calculating}. The choice of AB-magnitude system or Vega-system is specified by selecting the appropriate radio button. If a starburst model is to be used, clicking on the Starburst99 radio button will allow the age and metallicity of the model galaxy spectra to be chosen from drop down menus. Either an instantaneous or continuous star formation model may be chosen by selecting the appropriate radio button. The relevant spectra is then selected from the catalogue. The redshift to place the model galaxy at is input into the text field. It is now possible to calculate the colour of the model galaxy, without any absorption, by simply hitting {\it Calculate Colour}. The colour will be printed to the chosen output file. Alternatively, the user may upload their own galaxy spectra by clicking on the appropriate radio button. The format for the spectra is described in Section \ref{SEC:calculating}.
\newline\newline
To investigate the effects of intergalactic absorption, the user may choose whether to use the redshift distribution of Meiksin (2006) (Eq. \eqref{MeiksinDistribution}) or Inoue \& Iwata (2008) (Eq. \eqref{InoueDistribution}) for the Lyman Limit Systems (LLSs) by selecting the appropriate radio button. The parameters for these distributions may be changed by clicking on the appropriate tab and entering the desired values into the text fields. The opacity values are taken from (Eq. \eqref{TauDistribution}), with variable parameter $\beta$. If either of these distributions are selected, the program will perform Monte Carlo simulations for the number of lines-of-sight (realisations) defined, by randomly drawing opacity and redshift values for the LLSs from the chosen distribution. Alternatively, there is an option to use custom defined properties of Lyman Limit Systems from an input file, as described in Section \ref{SEC:methodTransmission}. If any of these methods are chosen, the contribution to the absorption by the Ly$\alpha$ forest will be included according to the model of Meiksin (2006) and the contribution by the optically thin IGM according to Eq. \eqref{diffuseIGM} for which the normalisation constant $A$ may be chosen. The resolution for calculating the transmission may also be chosen.
\newline\newline
Several options are available for the output. By checking or un-checking the relevant box, it is possible to either print the colours for each realisation separately or print the average over all lines of sight to the user-defined output file. There is also an option to print to file the cumulative distribution of colours. If absorption is included, the program will also print to file; the average transmission curve (see Section \ref{SEC:resultsTransmission}), the number and properties of the LLSs used for each line of sight and the average IGM k-correction values for each filter (as calculated from Eq. \eqref{kCorrection}), which may be useful for interpreting the results. 

\section{Integration using Gaussian quadrature}\label{GaussianQuadrature}
When numerical integration is required, Gaussian quadrature has been used. This involves approximating the integral of a function $f(x)$ by 
\begin{equation}
\int_{b}^{a} W(x)f(x)dx = \sum_{j=1}^{N}w_jf(x_j)
\end{equation}
where $W(x)$ is the weight function, $w_j$ are the weights and $x_j$ are abscissa points. The function is evaluated at each $x_j$ value. The code uses the weight function $W(x)=1$, corresponding to the {\it Gauss-Legendre} case. The weights in this case are given by
\begin{equation}
w_j = \frac{2}{(1-x_j^2)[P'_N(x_j)]^2}
\end{equation}
where $P'_N$ are the Legendre Polynomials. A routine to compute the integral between $x_1$ and $x_2$ was adapted from {\it Numerical Recipes in C} (Press et al. 1992), and translated into Java. The transmission functions $T(\lambda)$, the galaxy spectra $f_{\lambda}$ and transmission functions $\exp(-\tau)$ were all evaluated at the same absicissa points using linear interpolation. This method was tested using a simple Simpson's rule routine to verify that both methods agreed.

\end{document}